\documentclass[english,prl, twocolumn, superscriptaddress]{revtex4}
\usepackage[T1]{fontenc}
\usepackage[latin9]{inputenc}
\usepackage{xcolor}
\usepackage{babel}
\usepackage{amsmath}
\usepackage{amssymb}
\usepackage{graphicx}
\usepackage{setspace}
\usepackage[unicode=true,pdfusetitle,
 bookmarks=true,bookmarksnumbered=false,bookmarksopen=false,
 breaklinks=false,pdfborder={0 0 1},backref=false,colorlinks=false]
 {hyperref}

\makeatletter
\@ifundefined{textcolor}{}
{%
 \definecolor{BLACK}{gray}{0}
 \definecolor{WHITE}{gray}{1}
 \definecolor{RED}{rgb}{1,0,0}
 \definecolor{GREEN}{rgb}{0,1,0}
 \definecolor{BLUE}{rgb}{0,0,1}
 \definecolor{CYAN}{cmyk}{1,0,0,0}
 \definecolor{MAGENTA}{cmyk}{0,1,0,0}
 \definecolor{YELLOW}{cmyk}{0,0,1,0}
}

\makeatother

\begin{document}
\title{\noindent Quantum fluxes at the inner horizon of a spinning black
hole}
\author{Noa Zilberman}
\email{noazilber@campus.technion.ac.il}
\affiliation{Department of Physics, Technion, Haifa 32000, Israel}
\author{Marc Casals}
\email{marc.casals@uni-leipzig.de}
\email{marc.casals@ucd.ie}
\email{mcasals@cbpf.br}
\affiliation{Institut f\"ur Theoretische Physik, Universit\"at Leipzig, Br\"uderstrasse 16, Leipzig 04103, Germany}
\affiliation{Centro Brasileiro de Pesquisas F\'isicas (CBPF), Rio de Janeiro, CEP 22290-180, Brazil}
\affiliation{School of Mathematics and Statistics, University College Dublin, Belfield, Dublin 4, D04 V1W8, Ireland}
\affiliation{Laboratoire Univers et Th\'eories, Observatoire de Paris, CNRS, Universit\'e PSL,
Universit\'e de Paris, 92190 Meudon, France}
\author{Amos Ori}
\email{amos@physics.technion.ac.il}
\affiliation{Department of Physics, Technion, Haifa 32000, Israel}
\author{Adrian C. Ottewill}
\email{adrian.ottewill@ucd.ie}
\affiliation{School of Mathematics and Statistics, University College Dublin, Belfield, Dublin 4, D04 V1W8, Ireland}

\date{\today}
\begin{abstract}
Rotating or charged classical black holes in isolation possess a special
surface in their interior, the \emph{Cauchy horizon}, beyond which
the evolution of spacetime (based on the equations of General Relativity)
ceases to be deterministic. In this work, we study the effect of a
quantum massless scalar field on the Cauchy horizon inside a rotating
(Kerr) black hole that  is evaporating via the emission of Hawking
radiation (corresponding to the field being in the Unruh state). We
calculate the flux components (in Eddington coordinates) of the renormalized
stress-energy tensor of the field on the Cauchy horizon, as functions
of the black hole spin and of the polar angle. We find that these
flux components are generically nonvanishing. Furthermore, we find
that the flux components change sign as these parameters vary. The
signs of the fluxes are important, as they provide an indication of
whether the Cauchy horizon expands or crushes (when backreaction is
taken into account). Regardless of these signs, our results imply
that the flux components generically diverge on  the Cauchy horizon
when expressed in coordinates which are regular there. This is the
first time that irregularity of the Cauchy horizon under a semiclassical
effect is conclusively  shown for (four-dimensional) spinning black
holes. 
\end{abstract}
\maketitle

\paragraph*{Introduction.}

The simplest  spacetime solutions describing classical  spinning
or charged black holes (BHs) reveal nontrivial spacetime structures,
in which the  geometry connects through an inner horizon (IH) to
another external universe \citep{Carter:1966,GravesBrill:1960}. But
does such a smoothly-traversable passage really exist inside a physically-realistic
 spinning BH? 

Already classically, it is known \citep{Ori:1992,DafermosLuk:2017}
that introducing  various perturbing fields on a spinning (Kerr)
BH background leads to  formation of a weak \citep{Tipler,Ori:2000}
null curvature singularity along the otherwise regular Cauchy horizon
(CH) -- the ingoing section of the IH  (see also \citep{BradyDrozMorsnik:1998,Ori:1999}).
With these classical results established, it is interesting to extend
the study to the effect  of \emph{quantum perturbations} within the
semiclassical theory. It has been widely anticipated \citep{Hiscock:1976,BirrellDavies:1978,Hiscock:1980,OttewillWinstanley:2000},
yet still inconclusive, that semiclassical effects would diverge at
the CH.  Such a divergence, if indeed it occurs, may drastically
affect the internal BH geometry, potentially preventing the IH traversability.
Clarifying this issue requires the computation of $\left\langle T_{\alpha\beta}\right\rangle _{\text{ren}}$,
the \emph{renormalized stress-energy tensor }(RSET), on BH interiors.
However, this involves various   challenges.  

The RSET
\emph{flux components}, $\left\langle T_{uu}\right\rangle _{\text{ren}}$
and $\left\langle T_{vv}\right\rangle _{\text{ren}}$ ($u,v$ being the Eddington coordinates, introduced later), are of particular
interest, as they may crucially modify (through backreaction) the
internal geometry of the BH -- especially at the CH vicinity
(as discussed in Ref.~\citep{FluxesRN:2020}, in the analogous spherical charged case). A nonvanishing $\left\langle T_{vv}\right\rangle _{\text{ren}}$
at the CH implies a divergence of the RSET  there \footnote{\label{fn:regularity}Nonvanishing $\left\langle T_{vv}\right\rangle _{\text{ren}}$ ($\left\langle T_{uu}\right\rangle _{\text{ren}}$)
 at the CH (EH)   implies divergence of the  RSET  in the corresponding,
regular, Kruskal coordinates.}. Furthermore, the signs of $\left\langle T_{vv}\right\rangle _{\text{ren}}$
and $\left\langle T_{uu}\right\rangle _{\text{ren}}$ might determine the nature of their accumulating
backreaction effect on the near-CH geometry (see Eq. (15) in Ref.~\citep{FluxesRN:2020}, whose generalization to Kerr is underway).  With a negative (positive) $\left\langle T_{vv}\right\rangle _{\text{ren}}$, an infalling observer should experience abrupt expansion (contraction). In addition, preliminary hints suggest that a positive $\left\langle T_{uu}\right\rangle _{\text{ren}}$
may shrink the CH toward zero size, while a negative $\left\langle T_{uu}\right\rangle _{\text{ren}}$
may expand it, potentially retaining its traversability.

The flux components $\left\langle T_{vv}\right\rangle _{\text{ren}}$
and $\left\langle T_{uu}\right\rangle _{\text{ren}}$ were recently
computed \citep{FluxesRN:2020} at the CH of a spherical charged (Reissner-Nordstr\"om,
RN) BH,  using point splitting \citep{Christensen:1976} -- and were found to be either positive or negative, depending
on the BH's charge-to-mass ratio. (See also \citep{FluxesRNext:2021,Hollands:2020cqg,Hollands:2020prd}.)

In this paper we address the same problem as in Ref.~\citep{FluxesRN:2020},
but this time in the Kerr geometry. This is obviously the most realistic
BH canonical solution, as astrophysical BHs are known to be spinning.
We shall explore the behavior of the semiclassical flux components
$\left\langle T_{uu}\right\rangle _{\text{ren}}$ and $\left\langle T_{vv}\right\rangle _{\text{ren}}$
at the Kerr CH (in the Unruh state, corresponding to an evaporating
BH) -- both on and off the pole ($\theta=0$). We
shall demonstrate that these fluxes can be positive
or negative at the CH, depending on the BH spin parameter and the
polar angle. This constitutes  a novel quantitative step towards
settling the issue of IH traversability for spinning BHs.

To regularize the (naively diverging) semiclassical fluxes, we
employ the method of subtracting another quantum state, thereby
curing the divergence (see \citep{Candelas:1980,ChristensenFulling:1977}; this method was also used recently in \citep{Hollands:2020cqg} for spherical BH interiors).
  Here we apply it to the Kerr CH, using a special
quantum state (also resembling \citep{Taylor:2019}) designed
for that purpose. Constructing this state will involve an excursion into the "negative-mass universe" (described below).

\paragraph*{Preliminaries.}

The Kerr geometry, representing a spinning vacuum BH of mass $M$
and angular momentum $aM$,  is described by the line element 

\begin{align}
\text{d}s^{2}=-\left(1-\frac{2Mr}{\rho^{2}}\right)\text{d}t^{2}+\frac{\rho^{2}}{\Delta}\text{d}r^{2}+\rho^{2}\text{d}\theta^{2}+\label{eq:Kerr_metric}\\
\left(r^{2}+a^{2}+\frac{2Mra^{2}}{\rho^{2}}\sin^{2}\theta\right)\sin^{2}\theta\text{d}\varphi^{2}-\frac{4Mra}{\rho^{2}}\sin^{2}\theta\text{d}\varphi\text{d}t\nonumber 
\end{align}
where $\rho^{2}\equiv r^{2}+a^{2}\cos^{2}\theta$ and $\Delta\equiv r^{2}-2Mr+a^{2}$.
The two solutions of the equation $g^{rr}=0$ (i.e. $\Delta=0$)
yield an event horizon (EH) at $r=r_{+}$ and an IH at $r=r_{-}$,
where $r_{\pm}\equiv M\pm\sqrt{M^{2}-a^{2}}$. 

\begin{figure}[h!]
\centering \includegraphics[width=8cm]{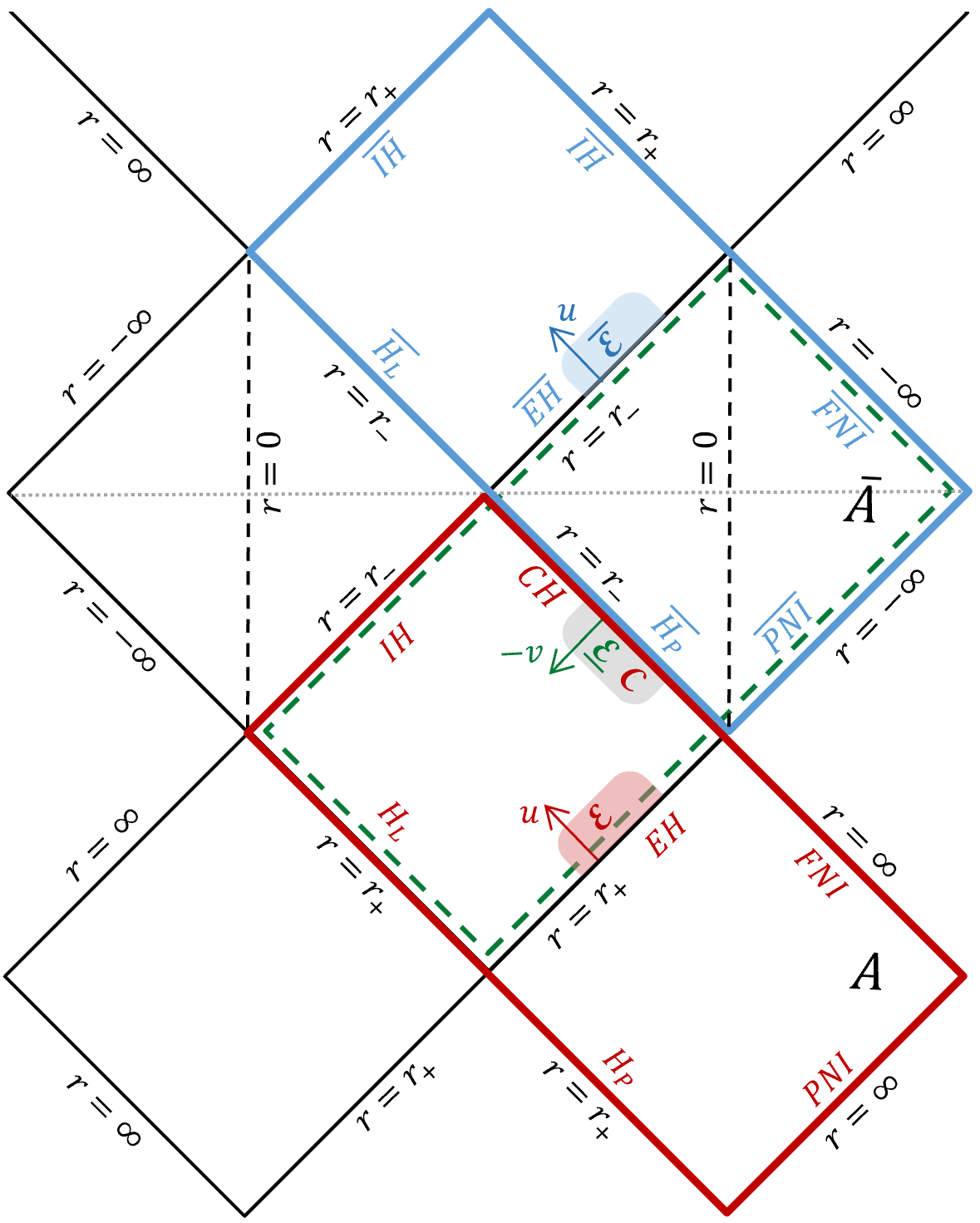}\caption{Penrose diagram of (part of) the analytically-extended
Kerr geometry. The two types of external universes are marked by
$A$ (\textquotedblleft usual\textquotedblright{} universe) and $\overline{A}$
(\textquotedblleft negative-mass\textquotedblright{} universe).  \emph{EH}
(\emph{IH}) marks the ordinary event (inner) horizon, whereas $\overline{EH}$
($\overline{IH}$) is  the event (inner) horizon of the analogous
$\overline{A}$-universe BH.  The standard Unruh-state domain is
bounded by the red frame. The domain for the $\overline{U}$ state
(associated with $\overline{A}$) is framed in blue, and its time-reversal
image (the $\underline{U}$-state domain) is the region bounded by
the green dashed frame. $\mathcal{E}$ (red shaded area) denotes
the internal near-EH region, $\mathcal{\overline{E}}$ (blue shaded
area) \textcolor{blue}{} its near-$\overline{EH}$ counterpart,
and $\mathcal{\underline{E}}$ (grey shaded area) \textcolor{blue}{}
the time-reversal image of $\mathcal{\overline{E}}$. $\mathcal{\underline{E}}$
coincides with $\mathcal{C}$, which denotes the CH vicinity.\textcolor{violet}{}}
\label{fig:penrose}
\end{figure}

The ingoing IH section marked \emph{``CH''}  in Fig.~\ref{fig:penrose}
is a CH with respect to initial data specified in the external universe
$A$. This null hypersurface plays a crucial role in the causal structure
of the BH. 

We consider  a minimally-coupled massless scalar field $\Phi\left(x\right)$,
satisfying $\square\Phi\equiv g^{\mu\nu}\Phi_{;\mu\nu}=0$.  This
field equation is separable in Kerr \citep{Carter:1968,Teukolsky:1972},
allowing solutions of the form
\begin{equation}
\Phi_{\omega lm}\left(t,r,\theta,\varphi\right)=\text{const}\cdot\frac{\psi_{\omega lm}\left(r\right)e^{im\varphi-i\omega t}}{\sqrt{r^{2}+a^{2}}}S_{lm}^{\omega}\left(\theta\right),\label{eq:decomKerr}
\end{equation}
where $S_{lm}^{\omega}\left(\theta\right)$ is the \emph{spheroidal}
\emph{wavefunction }\citep{Flammer}\emph{} and $\psi_{\omega lm}\left(r\right)$
is the  \emph{radial function}, satisfying

\begin{equation}
\frac{\text{d}^{2}\psi_{\omega lm}}{\text{d}r_{*}^{2}}+V_{\omega lm}\left(r\right)\psi_{\omega lm}=0\,.\label{eq:radKerr}
\end{equation}
Here $r_{*}$ is the tortoise coordinate satisfying $\text{d}r/\text{d}r_{*}=\Delta/\left(r^{2}+a^{2}\right)$.
   The effective potential $V_{\omega lm}\left(r\right)$ is
explicitly given in the Supplemental Material \citep{Sup} and satisfies
\begin{equation}
V_{\omega lm}\simeq\omega_{\pm}^{2},\,\,\,\,\,\,r\to r_{\pm}\,,\label{eq:asymPot_int}
\end{equation}
where $\omega_{\pm}\equiv\omega-m\Omega_{\pm}$ and $\Omega_{\pm}\equiv a/\left(2Mr_{\pm}\right)$. The parameter $\Omega_\pm$
 is also used to define azimuthal coordinates $\varphi_{\pm}\equiv\varphi-\Omega_{\pm}t$
regular at $r=r_{\pm}$, respectively.

We consider solutions to the radial equation in the BH interior, $\psi_{\omega lm}^{\text{int}}\left(r\right)$,
emerging as free waves from the EH:

\begin{equation}
\psi_{\omega lm}^{\text{int}}\simeq e^{-i\omega_{+}r_{*}},\,\,\,\,\,r\to r_{+}\,.\label{eq:psi_EH}
\end{equation}

\noindent From Eq.~(\ref{eq:asymPot_int}),  $\psi_{\omega lm}^{\text{int}}\left(r\right)$
admits the free near-IH asymptotic form:
\begin{equation}
\psi_{\omega lm}^{\text{int}}\simeq A_{\omega lm}e^{i\omega_{-}r_{*}}+B_{\omega lm}e^{-i\omega_{-}r_{*}},\,\,\,\,\,\,r\to r_{-}\,,\label{eq:psi_nearIH_asym}
\end{equation}
with constant coefficients $A_{\omega lm}$ and $B_{\omega lm}$.


We introduce the Eddington coordinates in the BH interior,  $u=r_{*}-t$
and $v=r_{*}+t$. The computation of the flux
components $\left\langle T_{uu}\right\rangle _{\text{ren}}$ and $\left\langle T_{vv}\right\rangle _{\text{ren}}$
at the CH is at the heart of this paper.

\paragraph*{The Unruh state and its bare mode contribution.}

It is particularly meaningful to compute the flux components in the
physically realistic Unruh state \citep{Unruh:1976} (denoted hereafter
by a superscript $U$).  This quantum state is defined by initial
conditions along the  two null hypersurfaces PNI and $H_{P}\cup H_{L}$,
where PNI (\emph{past null infinity}), $H_{P}$ (\emph{past horizon})
and $H_{L}$ (\emph{left horizon}) are shown (in red) in Fig.~\ref{fig:penrose}.
It then evolves according to the field equation throughout its future
domain of dependence, enclosed by the red frame in Fig.~\ref{fig:penrose}.

The Unruh state is thus regular throughout the interior of the red
frame, and in particular at the EH, which implies  $\left\langle T_{uu}\right\rangle _{\text{ren}}^{U}=0$
there [32].

Each mode contributes individually to the  fluxes.  In Appendix
B of Ref.~\citep{HTPFinKerr:2022}, we constructed the ``bare''
mode-sum expression (namely, prior to  regularization) for
the Unruh fluxes, $\left\langle T_{uu}\right\rangle _{\text{bare}}^{U}$
and $\left\langle T_{vv}\right\rangle _{\text{bare}}^{U}$, evaluated
at  the CH and EH. 

To express the results compactly, we hereby introduce the summation/integration
operator
\[
\hat{\sum}_{\pm}\left(\dots\right)\equiv\hbar\int_{0}^{\infty}\sum_{l=0}^{\infty}\sum_{m=-l}^{l}\frac{\left[S_{lm}^{\omega}\left(\theta\right)\right]^{2}}{8\pi^{2}\left(r_{\pm}^{2}+a^{2}\right)}\left(\dots\right)\text{d}\omega\,.
\]

Hereafter,  a superscript ``-'' (``+'') in $T_{uu}^{\pm}$
or $T_{vv}^{\pm}$  denotes the CH (EH)  limit, particularly referring
to evaluation in the shaded region marked $\mathcal{\mathcal{C}}$
($\mathcal{E})$ in Fig.~\ref{fig:penrose}, taking the $r\to r_{-}$
($r_{+}$) limit therein  \footnote{This $\pm$ superscript also indicates the  coordinate system in
use, being $\left(u,v,\theta,\varphi_{\pm}\right)$ at $r_{\pm}$.
}.

We concentrate now on $\left\langle T_{vv}^{-}\right\rangle _{\text{ren}}^{U}$,
leaving $\left\langle T_{uu}^{-}\right\rangle _{\text{ren}}^{U}$
to be treated afterwards. We may write $\left\langle T_{vv}^{-}\right\rangle _{\text{bare}}^{U}=\hat{\sum}_{-}E_{vv\left(\omega lm\right)}^{U-}$,
with $E_{vv\left(\omega lm\right)}^{U-}$  given by (see Eq.~(B49)
in  \citep{HTPFinKerr:2022}):
\begin{align}
E_{vv\left(\omega lm\right)}^{U-}=\frac{\omega_{-}^{2}}{\omega_{+}}\left[\coth\hat{\omega}_{+}\bigl(\left|A_{\omega lm}\right|^{2}+\left|\rho_{\omega lm}^{\text{up}}\right|^{2}\left|B_{\omega lm}\right|^{2}\bigr)+\right.\label{eq:EvvU-}\\
\left.2\text{cosech}\,\hat{\omega}_{+}\,\Re\left(\rho_{\omega lm}^{\text{up}}A_{\omega lm}B_{\omega lm}\right)+\bigl(1-\left|\rho_{\omega lm}^{\text{up}}\right|^{2}\bigr)\left|B_{\omega lm}\right|^{2}\right]\nonumber 
\end{align}
where  $\hat{\omega}_{\pm}\equiv\pi\omega_{\pm}/\kappa_{\pm}$,
$\kappa_{\pm}=\left(r_{+}-r_{-}\right)/4Mr_{\pm}$ and $\rho_{\omega lm}^{\text{up}}$
is the \emph{up }mode reflection coefficient (see e.g. \citep{HTPFinKerr:2022}).

Later we shall also need $\left\langle T_{uu}^{+}\right\rangle _{\text{bare}}^{U}$,
given by (see Eq.~(B45) in  \citep{HTPFinKerr:2022})

\begin{equation}
\left\langle T_{uu}^{+}\right\rangle _{\text{bare}}^{U}=\hat{\sum}_{+}E_{uu\left(\omega lm\right)}^{U+}\,,\,\,\,E_{uu\left(\omega lm\right)}^{U+}=\omega_{+}\coth\hat{\omega}_{+}\,.\label{eq:TuuU+}
\end{equation}

\paragraph*{The negative-mass universe.}

The entire construction given above for the Unruh state was based
in the red frame, corresponding to the ``usual'' asymptotically
flat universe $A$. We now  shift to the other asymptotically
flat universe, marked by $\overline{A}$  in Fig.~\ref{fig:penrose},
and attempt to use   it as a basis for constructing an analogous
Unruh-like state. 

In this universe $\overline{A}$, the value of $r$ steadily decreases
going outside,  and it approaches $r\to-\infty$ at spacelike (and
null) infinity, rather than  $+\infty$. Wishing to treat $\overline{A}$
 as we treat ``conventional'' asymptotically-flat universes (like
$A$),  we  transform to a new radial coordinate $\overline{r}\equiv-r$.
The new metric  then takes exactly the same form as the original
metric (\ref{eq:Kerr_metric}), with $r$ replaced by $\overline{r}$
and $M$ by the \emph{negative} mass parameter $\overline{M}\equiv-M$.
A far observer ($|r|\gg M)$ in this external universe will be \emph{gravitationally
repelled} by the central object.  We shall therefore refer to 
$\overline{A}$ as the \emph{negative-mass universe}. We  denote
the future-  (past-) null infinity of $\overline{A}$ by $\overline{FNI}$
 ($\overline{PNI}$), see Fig.~\ref{fig:penrose}.

This universe $\overline{A}$ has two  important features  distinguishing
it from the standard universe $A$: (i) The ``ring singularity'',
 located at $r=0$ (and $\theta=\pi/2$), and (ii) the presence of
closed timelike curves (CTCs). We shall return to address these aspects
later on. Nevertheless, the negative-mass universe shares various
properties with  $A$. Most remarkably, it admits its own \emph{black
hole}, whose event horizon is the null curve denoted by $\overline{EH}$
(see Fig.~\ref{fig:penrose}): All points to the bottom-right of
this null hypersurface can signal to $\overline{FNI}$ (along causal curves), whereas all points to its top-left
cannot. Furthermore, the inverse metric component  $g^{\overline{r}\,\overline{r}}$
changes  sign at  two $\overline{r}$ values   given by the standard
formula $\overline{r_{\pm}}=\overline{M}\pm(\,\overline{M}^{2}-a^{2})^{1/2}$.
(The coordinate $\overline{r}$ is therefore timelike at $\overline{r_{-}}<\overline{r}<\overline{r_{+}}$
and spacelike elsewhere.) Notice  that $\overline{r_{\pm}}=-r_{\mp}$.
Summarizing, the $\overline{A}$-universe event (inner) horizon, denoted
 $\overline{EH}$ ($\overline{IH}$) in Fig.~\ref{fig:penrose},
 is located at $\overline{r}=\overline{r_{+}}$ ($\overline{r_{-}}$),
which corresponds to $r=r_{-}$ ($r_{+}$).

\paragraph*{The $\overline{U}$ and $\underline{U}$ states.}

The entire construction of the Unruh state may be repeated analogously
in the blue frame (see Fig.~\ref{fig:penrose}).  That is, while
the original Unruh state is fed by initial conditions along the 
null hypersurfaces PNI and $H_{P}\cup H_{L}$, the new state, hereafter
denoted by $\overline{U}$,  is fed by fully analogous initial conditions
along the  corresponding  null hypersurfaces $\overline{PNI}$ (where
$\overline{r}\to\infty$) and $\overline{H_{P}}\cup\overline{H_{L}}$
(where $\overline{r}=\overline{r_{+}}$ ): bearing positive Eddington
frequencies  along $\overline{PNI}$, and positive Kruskal frequencies
\citep{Unruh:1976}  along $\overline{H_{P}}\cup\overline{H_{L}}$.
 It thus functions like the original Unruh state, but with respect
to the ``barred'', negative-mass, universe $\overline{A}$ (rather
than $A$).

The presence of CTCs in the domain $r_{-}>r>-\infty$ (as well as
a ring singularity at $r=0$) may challenge the construction of a
quantum state in the blue frame. Indeed, there is no well defined
Cauchy evolution for initial data specified at $\overline{PNI}$ and
$\overline{H_{P}}\cup\overline{H_{L}}$. Note, however, that the field
separability provides an alternative framework for defining the evolution:
One can decompose the initial data into separable field modes, and
then evolve each mode independently (by solving its radial equation).
The evolution of each  mode is well defined throughout the blue
frame. To see this, it is sufficient to note that the potential
$V_{\omega lm}\left(r\right)$ is regular at $r=0$ (indeed, on
the entire $r$-axis, see \textcolor{blue}{} \citep{Sup}). We
may use this modewise scheme to uniquely evolve the field modes
throughout the blue frame, and thereby  construct our $\overline{U}$
state. (Note that even in the ordinary Unruh state the computation
of the fluxes is usually done by summing/integrating over the individual
modes' contributions -- which can be done also for the
$\overline{U}$ state without obstacles.)

We now focus on the blue shaded region right above $\overline{EH}$
in Fig.~\ref{fig:penrose}, denoted $\mathcal{\overline{E}}$, which
is the ``barred'' counterpart of $\mathcal{E}$. We wish to compute
the $\overline{U}$-state $\left\langle T_{uu}\right\rangle _{\text{bare}}$
in this near-$\overline{EH}$ domain. The mode-sum computation (carried
out in  \citep{HTPFinKerr:2022})  that eventually led to Eq.~(\ref{eq:TuuU+}),
equally applies to the $\overline{U}$-state in the blue frame: One
just needs to replace $M$ by $-M$; and,  since  $\left\langle T_{uu}\right\rangle _{\text{bare}}$
is now evaluated at $\overline{EH}$ (rather than \emph{EH}), 
$r_{+}$ is replaced by $r_{-}$.  This results in changing $\Omega_{+}\mapsto\Omega_{-}$
and $\kappa_{+}\mapsto\kappa_{-}$ \footnote{The same considerations that led to defining  $\Omega_{+}\equiv a/2Mr_{+}$
in the original universe lead to defining  in the ``barred'' universe
$\overline{\Omega_{+}}\equiv a/2\overline{M}\overline{r_{+}}=a/2Mr_{-}=\Omega_{-}$.
A similar argument leads to $\overline{\kappa_{+}}=\kappa_{-}$. }, and, consequently, also $\omega_{+}\mapsto\omega_{-}$ and $\hat{\omega}_{+}\mapsto\hat{\omega}_{-}$.
The $U\mapsto\overline{U},\mathcal{E}\mapsto\mathcal{\overline{E}}$ counterpart
of Eq.~(\ref{eq:TuuU+}) therefore reads 
\begin{equation}
\langle T_{uu}^{(\mathcal{\overline{E}})}\rangle_{\text{bare}}^{\overline{U}}=\hat{\sum}_{-}E_{uu\left(\omega lm\right)}^{\overline{U}(\mathcal{\overline{E}})}\,,\,\,\,E_{uu\left(\omega lm\right)}^{\overline{U}(\mathcal{\overline{E}})}=\omega_{-}\coth\hat{\omega}_{-}\,.\label{eq:TuuU+bar}
\end{equation}
The superscript $(\mathcal{\overline{E}})$ marks the specific
location of evaluation. Moreover, note that the same regularity argument
that led to $\left\langle T_{uu}^{+}\right\rangle _{\text{ren}}^{U}=0$,
now implies $\langle T_{uu}^{(\mathcal{\overline{E}})}\rangle_{\text{ren}}^{\overline{U}}=0$
(since $\overline{EH}$ is enclosed by the blue frame).

Finally, we perform a time-reversal transformation of the ``barred''
universe and the $\overline{U}$ state based on it. This acts as mirroring
through the horizontal dotted line in Fig.~\ref{fig:penrose}, and
takes the blue frame to the dashed green frame therein, where we define
the $\underline{U}$ state as the time reversal of the $\overline{U}$
state. In particular, $\mathcal{\overline{E}}$ is  mapped to the
grey shaded region $\mathcal{\underline{E}}$, just below  \emph{CH}.
 This is  the main region of interest for our computation, since
it coincides with the near-CH domain $\mathcal{C}$, as seen in Fig.
\ref{fig:penrose}. This time reversal takes the $u$ direction in
$\mathcal{\overline{E}}$ to the $-v$ direction in  $\mathcal{\underline{E}}$
(as indicated by the blue arrow in $\mathcal{\overline{E}}$ which
is mapped to the green arrow in  $\mathcal{\underline{E}}$).  The
$\overline{U}$-state $T_{uu}$  in $\mathcal{\overline{E}}$ (given
in Eq.~(\ref{eq:TuuU+bar})) then matches the $\underline{U}$-state
$T_{vv}$ in  $\mathcal{\underline{E}}=\mathcal{C}$,  namely:
\begin{equation}
\left\langle T_{vv}^{-}\right\rangle _{\text{bare}}^{\underline{U}}=\hat{\sum}_{-}E_{vv\left(\omega lm\right)}^{\underline{U}-}\,,\,\,\,E_{vv\left(\omega lm\right)}^{\underline{U}-}=\omega_{-}\coth\hat{\omega}_{-}\,;\label{eq:TvvU_-}
\end{equation}
and, similarly,
\begin{equation}
\left\langle T_{vv}^{-}\right\rangle _{\text{ren}}^{\underline{U}}=\langle T_{uu}^{(\mathcal{\overline{E}})}\rangle_{\text{ren}}^{\overline{U}}=0\label{eq:Tvv0}
\end{equation}
(the rightmost equality was already established above). 

\paragraph*{Regularization. }

We now apply the  procedure of regularization  by subtracting 
the $\underline{U}$-state from the Unruh state, recalling that
the difference between the  bare mode-sums of the two states is regular,
and equals the difference between the renormalized quantities. That
is,
\[
\left\langle T_{vv}^{-}\right\rangle _{\text{ren}}^{U}-\left\langle T_{vv}^{-}\right\rangle _{\text{ren}}^{\underline{U}}=\hat{\sum}_{-}\left(E_{vv\left(\omega lm\right)}^{U-}-E_{vv\left(\omega lm\right)}^{\underline{U}-}\right)\,.
\]
Recalling Eqs.~(\ref{eq:TvvU_-},\ref{eq:Tvv0}), our final
expression for $\left\langle T_{vv}^{-}\right\rangle _{\text{ren}}^{U}$
is thus 
\begin{equation}
\left\langle T_{vv}^{-}\right\rangle _{\text{ren}}^{U}=\hat{\sum}_{-}\left(E_{vv\left(\omega lm\right)}^{U-}-\omega_{-}\coth\hat{\omega}_{-}\right)\,,\label{eq:TvvU-ren}
\end{equation}
where $E_{vv\left(\omega lm\right)}^{U-}$ was  specified in Eq.
(\ref{eq:EvvU-}), and, recall $\hat{\omega}_{\pm}\equiv\pi\omega_{\pm}/\kappa_{\pm}$.

Finally, we consider $\left\langle T_{uu}^{-}\right\rangle _{\text{ren}}^{U}$.
 In Eq.~(B51) in \citep{HTPFinKerr:2022} we found the difference:
\begin{align}
 & \left\langle T_{uu}^{-}\right\rangle _{\text{ren}}^{U}-\left\langle T_{vv}^{-}\right\rangle _{\text{ren}}^{U}=\label{eq:Tvv-Tuu}\\
 & \hat{\sum}_{-}\omega_{-}\left(\coth\hat{\omega}_{+}-1\right)\left(1-\left|\rho_{\omega lm}^{\text{up}}\right|^{2}\right)\,,\nonumber 
\end{align}
which, along with Eq.~(\ref{eq:TvvU-ren}), yields $\left\langle T_{uu}^{-}\right\rangle _{\text{ren}}^{U}$.

We have thus obtained  simple and useful expressions for $\left\langle T_{vv}^{-}\right\rangle _{\text{ren}}^{U}$
and $\left\langle T_{uu}^{-}\right\rangle _{\text{ren}}^{U}$. 

\paragraph*{Numerical results.}

Next we numerically compute $\left\langle T_{vv}^{-}\right\rangle _{\text{ren}}^{U}$
and $\left\langle T_{uu}^{-}\right\rangle _{\text{ren}}^{U}$ based
on Eqs.~(\ref{eq:TvvU-ren}),(\ref{eq:Tvv-Tuu}).

 We start at the pole, where only $m=0$ modes contribute
(since $S_{l,m\neq0}^{\omega}\left(\theta=0\right)$ vanishes), which
drastically simplifies the numerical application.  We numerically
compute $A_{\omega lm},B_{\omega lm},\rho_{\omega lm}^{\text{up}}$
and construct  the integrand in Eq.~(\ref{eq:TvvU-ren}) (see 
\citep{Sup} for  details).  We find that all divergences present
in $\left\langle T_{vv}^{-}\right\rangle _{\text{bare}}^{U}$   entirely disappear in its renormalized counterpart. This provides a crucial  test for our state-subtraction procedure.
In fact, the integrand converges exponentially in both $l$ and $\omega$.

\begin{figure}[h!]
\centering \includegraphics[width=8.7cm]{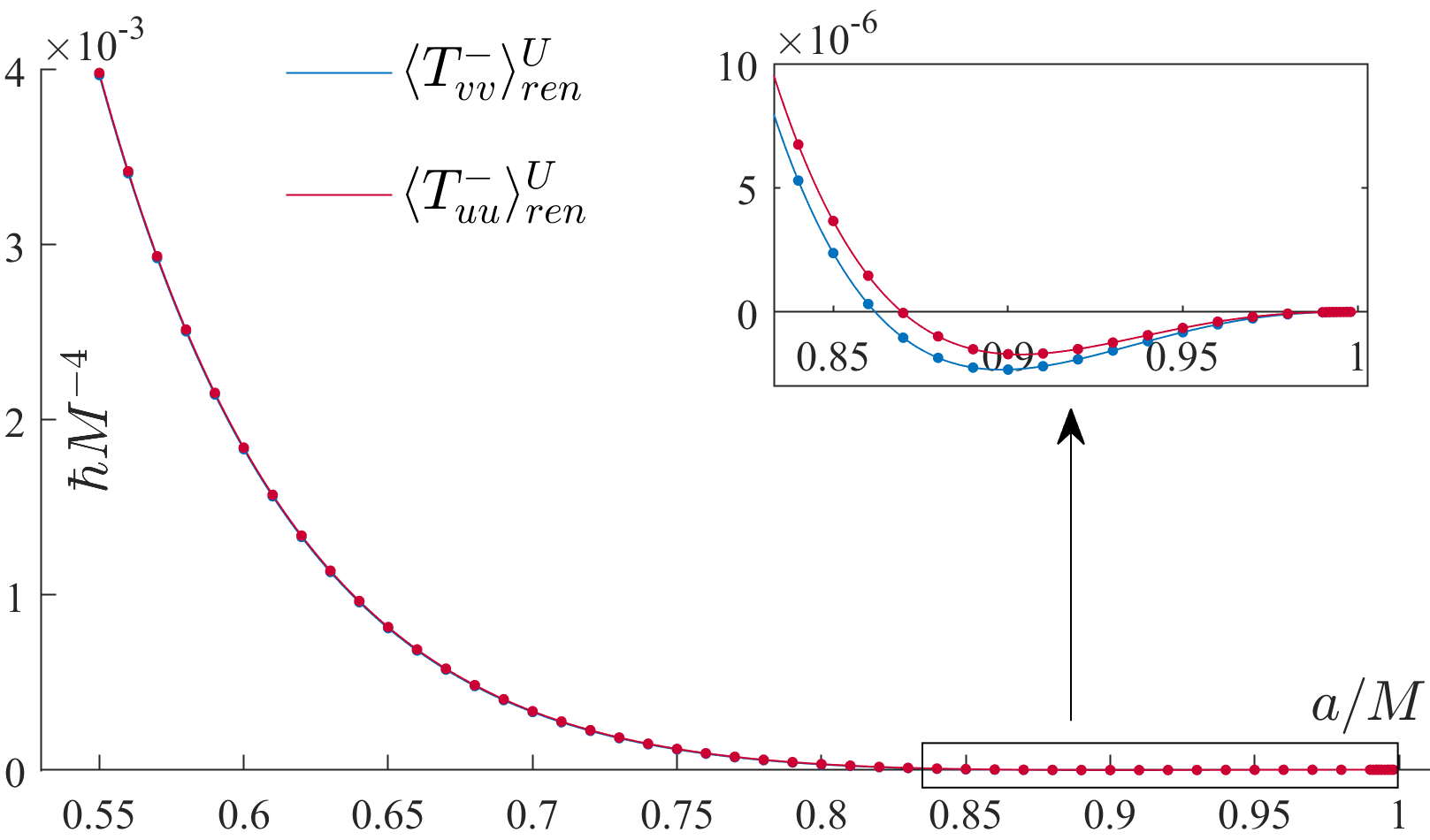}\caption{The polar  CH-limit Unruh fluxes  versus $a/M$, with the sign-flip
domain zoomed-in  at the top-right corner. The  dots were numerically
computed,  while the connecting lines are interpolated. Note that
wherever only the red is visible, it covers the blue.}
\label{fig:polar fluxes}
\end{figure}

Fig.~(\ref{fig:polar fluxes}) portrays  $\left\langle T_{vv}^{-}\right\rangle _{\text{ren}}^{U}$
and $\left\langle T_{uu}^{-}\right\rangle _{\text{ren}}^{U}$  at
the pole versus $a/M$.

It would be desirable to compare our results with those obtained by
other regularization methods. For the specific cases $a/M=0.8$ and $0.9$, we
computed \citep{future} the polar fluxes via point splitting \citep{Christensen:1976}
for various $r_{-}<r<r_{+}$ values.   We then evaluated 
the $r\to r_{-}$ limit of these fluxes (see  \citep{Sup}). We find
full agreement between the two methods (point splitting and state
subtraction). E.g., for $a/M=0.8$, both methods yield $\left\langle T_{vv}^{-}\right\rangle _{\text{ren}}^{U}\approx0.00003013\,\hbar M^{-4}$
and $\left\langle T_{uu}^{-}\right\rangle _{\text{ren}}^{U}\approx0.00003232\,\hbar M^{-4}$
\footnote{In the point-splitting (state-subtraction) method we obtain these
values with four (nine) significant figures.}. This excellent agreement  strongly corroborates our state-subtraction
method.

It is interesting to compare the features seen here to the analogous RN case \citep{FluxesRN:2020}.  The CH-limit
polar fluxes in Kerr, like  in RN (with $Q/M$ replacing $a/M$),
are increasingly positive  for smaller spin values. They decrease
with increasing $a/M$ and change their sign at some critical value
beyond which they are negative all the way to their decay at $a/M\to1$
(more details in \citep{Sup}). The critical sign-flip
values are smaller here compared to their RN counterparts \citep{FluxesRN:2020},
being  $a/M\approx0.862$ for $\left\langle T_{vv}^{-}\right\rangle _{\text{ren}}^{U}$
and $\approx0.870$ for $\left\langle T_{uu}^{-}\right\rangle _{\text{ren}}^{U}$.
 Moreover, numerically investigating the mentioned near-extremal
decay  versus the small parameter $\epsilon\equiv\sqrt{1-\left(a/M\right)^{2}}$
we obtain that, in full analogy with the RN case \citep{FluxesRNext:2021},
$\left\langle T_{vv}^{-}\right\rangle _{\text{ren}}^{U}\propto\epsilon^{4}$
and $\left\langle T_{uu}^{-}\right\rangle _{\text{ren}}^{U}\propto\epsilon^{5}$
(see \citep{Sup}). 

\begin{figure}[h!]
\centering \includegraphics[width=8.7cm]{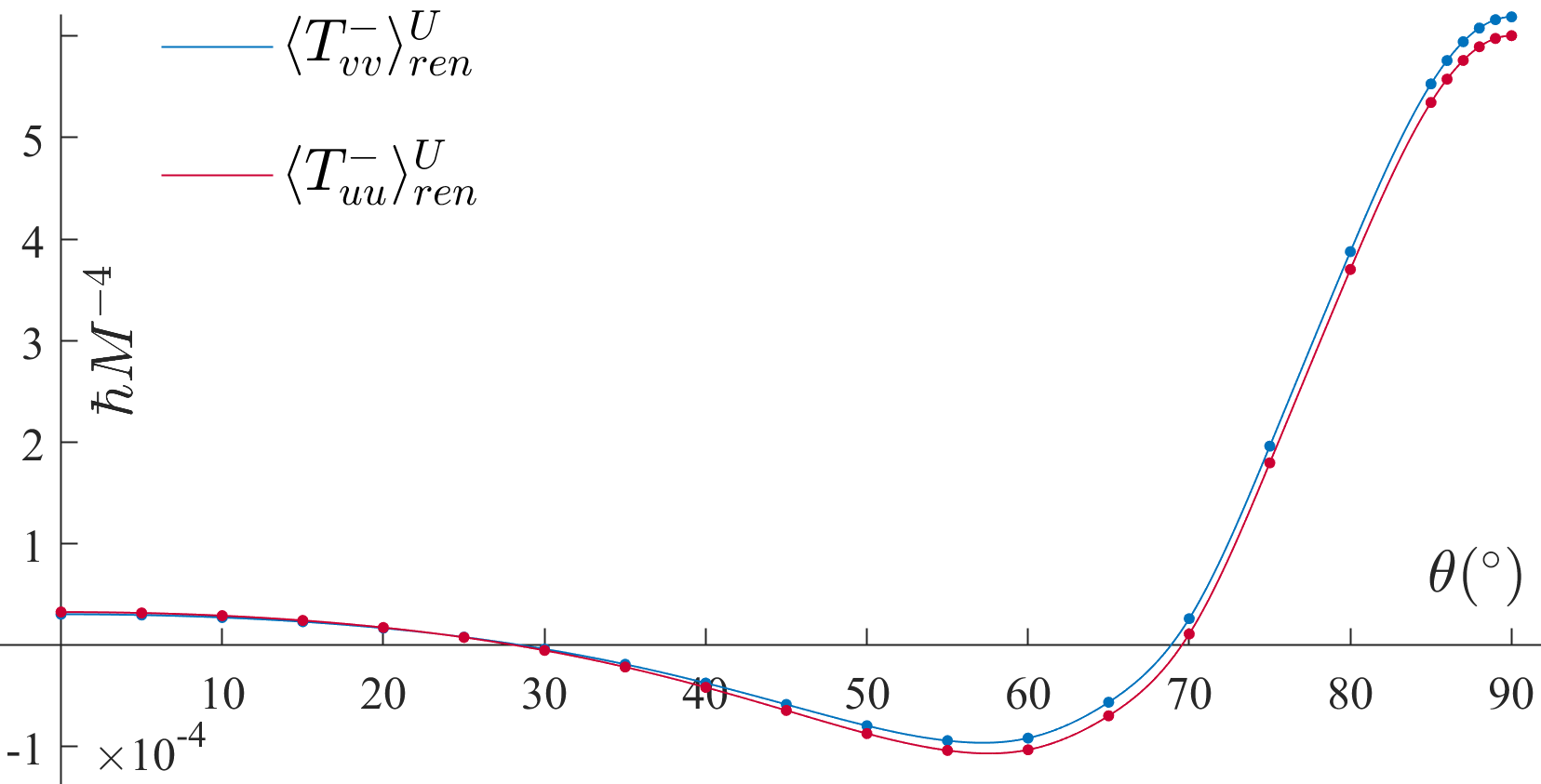}\caption{The CH-limit Unruh fluxes  versus  $\theta$,  for $a/M=0.8$.
The  dots were numerically computed and the connecting lines are
interpolated.}
\label{fig:off the pole}
\end{figure}

Next, we  compute the fluxes at other $\theta$ values, for $a/M=0.8$.
  We again find exponential convergence of the integrand  in both
$l$ and $\omega$ -- supporting the validity of our regularization
method off-pole as well. Fig.~\ref{fig:off the pole} displays our
results. Interestingly, the CH-limit fluxes change their sign
twice as a function of $\theta$ until they peak at the equator (around which they are symmetric).

\paragraph*{Conclusion.}

We computed the semiclassical Unruh-state fluxes $\left\langle T_{uu}\right\rangle _{\text{ren}}$
and $\left\langle T_{vv}\right\rangle _{\text{ren}}$ at the CH of
a spinning BH, using the state-subtraction method. We found generically
nonvanishing $\left\langle T_{vv}\right\rangle _{\text{ren}}$ at
the CH, implying the divergence of the RSET (and tidal forces) there. Furthermore, we found
that these fluxes may be positive or negative, depending on
$a/M$ and $\theta$. The sign of these fluxes may be crucial for the
nature of backreaction (see Introduction).

The quantum state $\underline{U}$ used for this subtraction is nonconventional
in several respects: First, it (partly) resides in the ``negative
mass'' asymptotic universe $\overline{A}$ in the analytically extended
Kerr geometry -- a spacetime region whose  existence in
a real spinning BH is at least highly questionable. Second, it is
a \emph{time-reversed} quantum state (with asymptotic boundary data
specified on the future rather than past null infinity of $\overline{A}$).
Third, this universe $\overline{A}$ contains  CTCs as well as a
naked ring singularity. Nevertheless, we use the
subtraction of this quantum state merely as a mathematical-computational tool and
it seems to work extremely well: First, it fully regularizes the
flux mode-sums. Furthermore, the subtracted mode-sum converges exponentially
fast, in both $\omega$ and $l$. Moreover, in two specific cases,
we compared the resultant flux values to those obtained by point
splitting, and found excellent quantitative agreement.

It will be important to extend this research to additional $a/M$
values (especially off-pole, where it would also be imperative to
compare to other methods)  -- and, even more importantly,
to the more realistic (quantum) electromagnetic field.

Computing the  semiclassical fluxes at the CH of a Kerr BH (and,
more importantly, determining their sign) is a major feat, but definitely
does not mark the end of this research:  Rather, these results open
a door to the study of  backreaction (via the semiclassical Einstein
equation) -- and the  resultant spacetime structure --
inside  a realistic, evaporating, spinning BH. We hope to further
explore this issue in a future work.

\begin{acknowledgments}
M.C. acknowledges partial financial support by CNPq (Brazil), process
number 314824/2020-0, and by the Scientific Council of the Paris Observatory
during a visit. A.O. and N.Z. were supported by the Israel Science
Foundation under Grant No. 600/18. N.Z. also acknowledges support
by the Israeli Planning and Budgeting Committee.
\end{acknowledgments}

\

\end{document}